# Electrically-Tunable High Curie Temperature Two-Dimensional Ferromagnetism in Van der Waals Layered Crystals

Hua Wang[1,†], Jingshan Qi[2,†,*], and Xiaofeng Qian[1,*]

[1]*Department of Materials Science and Engineering, Texas A&M University, College Station, Texas 77843, USA*

[2]*School of Physics and Electronic Engineering, Jiangsu Normal University, Xuzhou 221116, People's Republic of China*

†These authors contributed equally to this work.

*Authors to whom correspondence should be addressed: qijingshan@jsnu.edu.cn, feng@tamu.edu

## ABSTRACT

Identifying intrinsic low-dimensional ferromagnets with high transition temperature and electrically tunable magnetism is crucial for the development of miniaturized spintronics and magnetoelectrics. Recently long-range 2D ferromagnetism was observed in van der Waals crystals $CrI_3$ and $Cr_2Ge_2Te_6$, however their Curie temperature is significantly lowered when reducing down to monolayer/few layers. Herein, using renormalized spin-wave theory and first-principles electronic structure theory, we present a theoretical study of electrically tunable 2D ferromagnetism in van der Waals layered CrSBr and CrSeBr semiconductors with high Curie temperature of ~150K and sizable band gap. High transition temperature is attributed to strong anion-mediated superexchange interaction and a sizable spin-wave excitation gap due to large exchange and single-ion anisotropy. Remarkably, hole and electron doping can switch magnetization easy axis from in-plane to out-of-plane direction. These unique characteristics establish monolayer CrSBr and CrSeBr as promising platform for realizing 2D spintronics and magnetoelectrics such as 2D spin field effect transistor.

Achieving long-range magnetism at low dimensions and high temperature is of both fundamental and technological importance.[1-3] Particularly, intrinsic low-dimensional semiconducting ferromagnets with high Curie temperature $T_c$, large band gap, and high carrier mobility will help go beyond dilute magnetic semiconductor[1] and pave the way for the development of next-generation ultra-miniaturized, highly integrated spintronics and magneto-optoelectronics.[2,3] However, the coexistence of ferromagnetic (FM) and semiconducting characteristics in a single material is generally difficult,[3,4] whereas achieving long-range magnetic ordering in two-dimensional (2D) materials is even harder.

According to the Mermin-Wagner theorem, 2D FM/antiferromagnetic (AFM) order is prohibited by thermal fluctuations within *isotropic* Heisenberg model with continuous SU(2) symmetry.[5] *Finite magnetic anisotropy* such as exchange anisotropy and single-ion anisotropy becomes critical for establishing long-range magnetic order, *e.g.* in ultrathin metallic films decades ago.[6,7] Long-range 2D FM order was also observed recently in van der Waals (vdW) insulators $CrI_3$[8] ($T_c$ of ~45K) and $Cr_2Ge_2Te_6$[9] ($T_c$ of ~25K), while their $T_c$ is markedly lowered with decreasing layers. High $T_c$ semiconducting ferromagnets are thus highly desirable,[10] which could further impact 2D multiferroics.[11] Recently, vdW layered chromium sulfur bromide (CrSBr) has attracted attention as its bulk, first synthesized 50 years ago, was an AFM semiconductor.[12,13] Several recent theoretical studies[14-17] predicted that monolayer CrSBr and CrSeBr are FM semiconductor with much higher $T_c$ compared with $CrI_3$ and $Cr_2Ge_2Te_6$, ranging from ~300K to 160K using 2D Ising model,[15] 2D Heisenberg model without magnetic anisotropy,[17] and 2D Heisenberg model with single-ion anisotropy.[16]

Herein, based on first-principles density-functional theory (DFT),[18,19] we develop an anisotropic Heisenberg XYZ model including both single-ion anisotropy and exchange anisotropy and a spin-wave theory for monolayer CrSBr and CrSeBr, and find that monolayer CrSBr and CrSeBr have high $T_c$ of 168K and 150K, respectively. Our results show that the combination of large spin-wave excitation gap and exchange constants leads to high $T_c$. We also found that monolayer CrSBr possesses a remarkable electrically tunable magnetic ordering where the magnetization easy axis can be tuned from in-plane to out-of-plane by electrostatic doping. We provide a microscopic mechanism for the origin of high $T_c$ and electrically controllable magnetism in monolayer CrSBr, which not only offers design rules for high-temperature FM semiconductors, but also allows for realizing spin field effect transistor (spin FET) in monolayer 2D materials.[20]

Electronic structure calculations were carried out using DFT as implemented in the Vienna Ab initio Simulation Package (VASP)[21,22] with the projector-augmented wave method.[23] Electronic band structures were calculated using HSE06 and PBE exchange-correlation functional.[24] Hubbard U corrections were included in the DFT-PBE calculations with $U_{eff}=U-J=3$ eV to account for the correlation effect from 3$d$ transition metal.[25] For magnetic anisotropy, spin-orbit coupling (SOC) is taken into account at the full-relativistic level. More calculation details are included in supplementary material.

Bulk CrXBr (X=S, Se) are vdW layered crystals with *Pmmn* orthorhombic space group, and their monolayer has *Pmmm* space group (Figs. 1(a-d)). The optimized structure (see Table S1) agree with experiment. The cleavage energy for CrSBr and CrSeBr is only about ~0.3 J/m$^2$, less than 0.465 J/m$^2$ of graphene calculated using the same vdW correlation functional (see Fig. S1). Dynamical stability is verified by phonon dispersion (see Fig. S2) with the absence of imaginary modes. Thus, monolayer CrSBr and CrSeBr may be easily exfoliated from their bulk counterpart.[13] Magnetic property of monolayer CrXBr is largely dependent on the local environment of $Cr^{3+}$ which can be viewed as a distorted octahedron with each $Cr^{3+}$ surrounded by four $S^{2-}$ ($Se^{2-}$) and two $Br^-$, as illustrated in Fig. 1(e). In an ideal octahedron, crystal field splitting breaks five-fold degenerate $d$ orbitals into double-degenerate $e_g$ and three-fold degenerate $t_{2g}$ orbitals. The presence of two types of anions further reduces it to $C_{2v}$, which lifts doubly-degenerate $e_g$ to two $a_1$ orbitals and lifts triply-degenerate $t_{2g}$ to $b_1$, $b_2$, and $a_2$ orbitals. Because spin pairing energy $U_p$ for transition from parallel spins on two orbitals to antiparallel spins on a single orbital is greater than crystal field splitting energy $\Delta_c$, parallel high spin state and hence ferromagnetism should be favored.

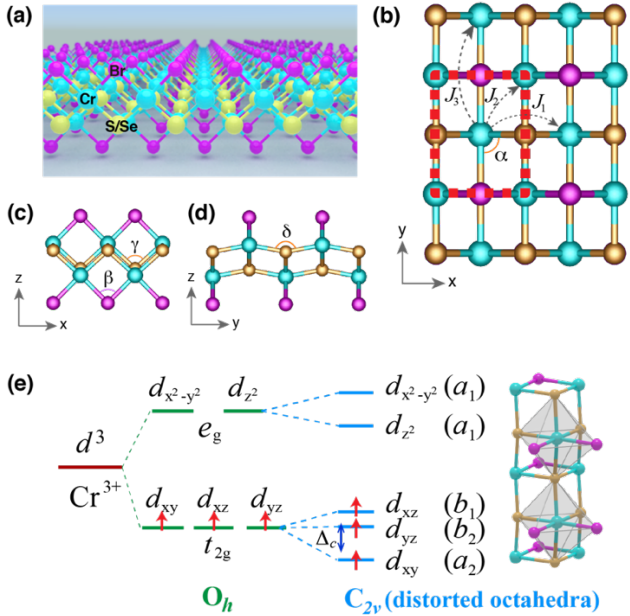

FIG. 1. Monolayer CrXBr and their structural properties. (a, b, c, d) Crystal structure of vdW layered crystals CrXBr (X=S, Se) in their monolayer form. (e) Crystal field splitting from an ideal octahedron with $O_h$ symmetry to a distorted octahedron with $C_{2v}$ symmetry.

To confirm the above analysis, we calculate the band structure with HSE06 hybrid functional. As shown in Figs. 2(a,b) and Figs. S3 and S4, monolayer CrSBr and CrSeBr exhibit highly anisotropic electronic structure and a band gap of 1.66 eV and 0.78 eV, respectively. This large anisotropy is evidenced by small effective hole mass of only ~0.11$m_0$ and large hole mobility of ~720 cm$^2$ V$^{-1}$ s$^{-1}$ (see Table S2).[26, 27] The HSE06 bandgap is higher than that from DFT-PBE calculations (0.2 and 0.8 eV for CrSBr and 0.1 eV for CrSeBr)[15-17] due to the inclusion of short-range screened non-local Fock exchange interaction. Next, we investigate ferromagnetism in monolayer CrXBr. Using a 2×2×1 supercell (Fig. S5), we first verified that FM ordering is more stable than AFM ordering. Spin density (Fig. S6) mainly comes from Cr with ~3$\mu_B$ per Cr, consistent with the high spin state of $Cr^{3+}$. In contrast, S/Se atoms carry small opposite spin moment and Br atoms carry smaller opposite spin moment as listed in Table S3 and Fig. S7 due to less (two) nearest neighboring Cr atoms for Br compared to those for S/Se. The FM coupling between Cr atoms is originated from superexchange interaction mediated by S (Se) and Br.[28-30] To illustrate this, we classify the linking geometry between $Cr^{3+}$ ions into two types. In type I, the neighboring octahedra have two common edge atoms S(Se) and Br where the Cr-S(Se)-Cr and Cr-Br-Cr angle is ~90º, *e.g.* α along *ab* diagonal, and β and γ along *a* axis (Figs. 1(b-c)). In type II, the neighboring octahedra along the *b* axis have one common corner S (Se) with a Cr-S(Se)-Cr angle δ of ~160º (Fig. 1(d)). According to Goodenough-Kanamori-Anderson rules,[30, 31] FM coupling is favored for 90º superexchange interaction between two magnetic ions with partially filled *d* shells, while AFM coupling is preferred for 180º superexchange interaction. As δ significantly deviates from 90º and 180º, the superexchange interaction along *b* exhibits competing FM and AFM coupling. In contrast, FM coupling along *a* axis and *ab* diagonal is strongly favored due to ~90º α, β and γ, establishing FM ground state.

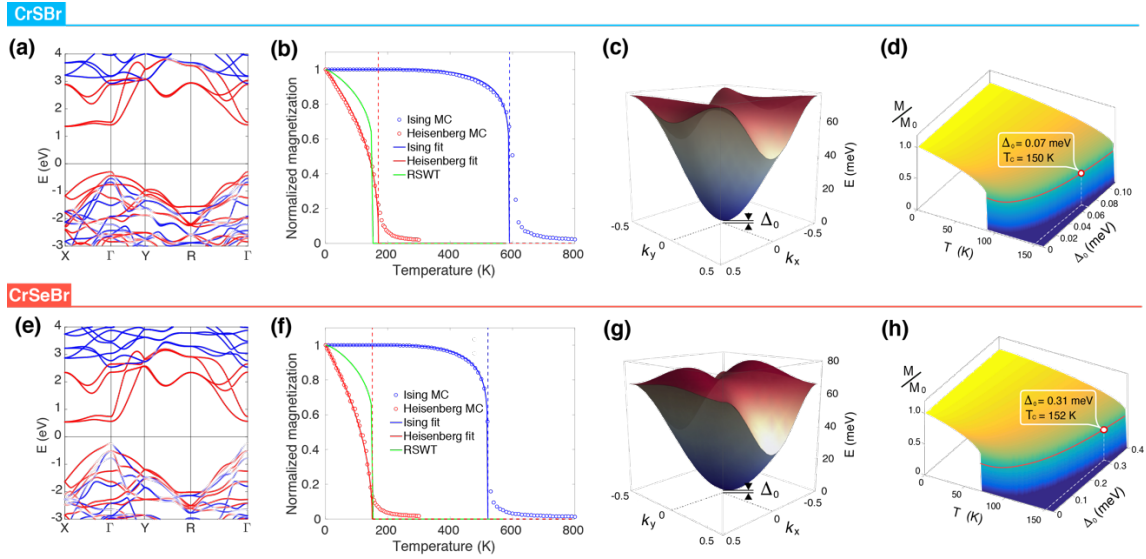

FIG. 2. Electronic structure and ferromagnetism in monolayer CrSBr and CrSeBr. (a, e) Electronic band structure with HSE06 functional. Red (blue) indicates spin majority (minority). (b, f) Temperature dependent normalized magnetization using different theoretical models, including Ising model, anisotropic Heisenberg model, and RSWT. (c, g) Magnon dispersion with spin-wave excitation gap $\Delta_0$ located at Γ. (d, h) Magnetic moment as function of temperature and spin-wave excitation gap. Two dots indicate the corresponding $T_c$ for monolayer CrSBr and CrSeBr.

Single-ion anisotropy and exchange anisotropy are two important sources that contribute to magnetocrystalline anisotropy energy (MAE) which is crucial for establishing long-range 2D ferromagnetic order. MAE values are listed in Table S5. It shows high magnetic anisotropy with easy axis along *a* (*i.e. x* direction), distinct from CrI$_3$[8] and Cr$_2$Ge$_2$Te$_6$[9] with out-of-plane easy axis. We then build the corresponding spin Hamiltonian with classical Heisenberg XYZ model,[32]

$$H = -\sum_{\langle ij \rangle} J_1 \vec{S}_i \cdot \vec{S}_j - \sum_{\langle\langle ij \rangle\rangle} J_2 \vec{S}_i \cdot \vec{S}_j - \sum_{\langle\langle\langle ij \rangle\rangle\rangle} J_3 \vec{S}_i \cdot \vec{S}_j - \sum_i D^x (S_i^x)^2 - \sum_i D^y (S_i^y)^2 - \sum_{\langle ij \rangle} \lambda_1^x S_i^x S_j^x - \sum_{\langle\langle ij \rangle\rangle} \lambda_2^x S_i^x S_j^x - \sum_{\langle\langle\langle ij \rangle\rangle\rangle} \lambda_3^x S_i^x S_j^x - \sum_{\langle ij \rangle} \lambda_1^y S_i^y S_j^y - \sum_{\langle\langle ij \rangle\rangle} \lambda_2^y S_i^y S_j^y - \sum_{\langle\langle\langle ij \rangle\rangle\rangle} \lambda_3^y S_i^y S_j^y$$

(1).

$J_{1,2,3}$ are isotropic Heisenberg exchange coupling constants for the first, second, and third nearest-neighbor (NN) spins, and $S_i$ is 3/2 for Cr$^{3+}$. $D^x$ and $D^y$ refer to single-ion anisotropy along *x* and *y*. The rest refer to the exchange anisotropy. We calculate these parameters by mapping magnetic configurations (Fig. S5) to the above Hamiltonian, listed in Table S5. With this Hamiltonian, we calculate the Curie temperature using Monte Carlo (MC) method implemented in the VAMPIRE code[33] using a supercell of 11,250 spins. The results are shown in Figs. 2(b,f) and listed in Table S5. The critical exponent is obtained by fitting magnetization to the Curie-Bloch equation, yielding $T_C$ of 168K and 150K and $\beta$ of 0.41 and 0.42 for monolayer CrSBr and CrSeBr, respectively. $T_c$ from Ising model is 590K and 520K, much higher than that from the Heisenberg model as expected. The calculated Curie temperatures are significantly higher than that of 2D CrI$_3$[8] (45K) and Cr$_2$Ge$_2$Te$_6$[9] (20K). Our present calculation is based on anisotropic Heisenberg XYZ model with up to the 3$^{rd}$ nearest neighbors, thus resulting in lower $T_c$ compared to the recent theoretical prediction with 2D Ising model[15] and 2D Heisenberg model without magnetic anisotropy.[17] The agreement with the prediction using 2D Heisenberg model with single-ion anisotropy[16] indicates the importance and large influence of magnetic anisotropy.

To better understand the physical origin of high $T_c$, we apply linear spin-wave approximation with Heisenberg XXZ model[32] and arrive at a second quantization representation using Holstein-Primakoff transformation,[34]

$$H_{\text{spin-wave}} = \sum_i \varepsilon_0 \, b_i^+ b_i - J_1 S \sum_{\langle ij \rangle} b_i^+ b_j - J_2 S \sum_{\langle\langle ij \rangle\rangle} b_i^+ b_j - J_3 S \sum_{\langle\langle\langle ij \rangle\rangle\rangle} b_i^+ b_j \quad (2)$$

The parameters were included in Table S6 of supplementary material. The resulted spin-wave excitation gap $\Delta_0$ at Γ point from Eq. 2 yields 0.07 meV for CrSBr and 0.31 meV for CrSeBr. The 2D magnon dispersion is shown in Figs. 2(a) and (d) for CrSBr and CrSeBr, respectively. We then estimate the Curie temperature $T_c$ based on renormalized spin-wave theory (RSWT),[32]

$$M(T) = S - \frac{1}{N_s N_k} \sum_n \sum_k \left[ \exp\left( \frac{M(T) E_n(k_x, k_y)}{S k_B T} \right) - 1 \right]^{-1} \quad (3)$$

where $n$, $N_s$ and $N_k$ refer to band index, number of spins per unit cell and number of $\mathbf{k}$ points sampled in the first Brillouin zone. For the RSWT calculations, a dense k-point sampling of 200×200×1 was used for the BZ integration. The calculated $T_C$ from RSWT are 150K and 152K for monolayer CrSBr and CrSeBr, respectively, in reasonable agreement with $T_C$ from MC simulations. In RSWT, bosonic operator is kept up to the fourth order followed by a mean field approximation, which essentially scales/renormalizes the hopping energy and spin-wave gap. As shown in Figs. 2(d,h), $T_c$ is strongly dependent on spin-wave excitation gap $\Delta_0$ which is further determined by exchange and single-ion anisotropy. The underlying physical mechanism of FM order in finite temperature is thus originated from magnetic anisotropy.[35]

High $T_c$ 2D ferromagnetism in monolayer CrSBr and CrSeBr opens up exciting opportunities. For example, externally controlled magnetism is highly desirable for magnetoelectrics. Recent works have shown that carrier doping can be introduced into monolayers by electric gating to control magnetism.[36-38] In addition, vdW gap in 2D magnetic layers induces giant tunneling magnetoresistance.[39, 40] Here we show that carrier doping can drastically change MAE in 2D CrSBr and switch the magnetization easy axis. As shown in Fig. 3(a), under a hole doping of $n > 4 \times 10^{12}$/cm$^2$, the out-of-plane magnetization becomes favored. In contrast, electron doping, though showing similar trend, has relatively weak influence on the magnetization crystalline anisotropy. Experimentally, it is possible to achieve carrier concentration of up to $10^{13}$-$10^{14}$/cm$^2$ in 2D materials, therefore carrier doping could be an effective strategy to control magnetization ordering in monolayer CrSBr.

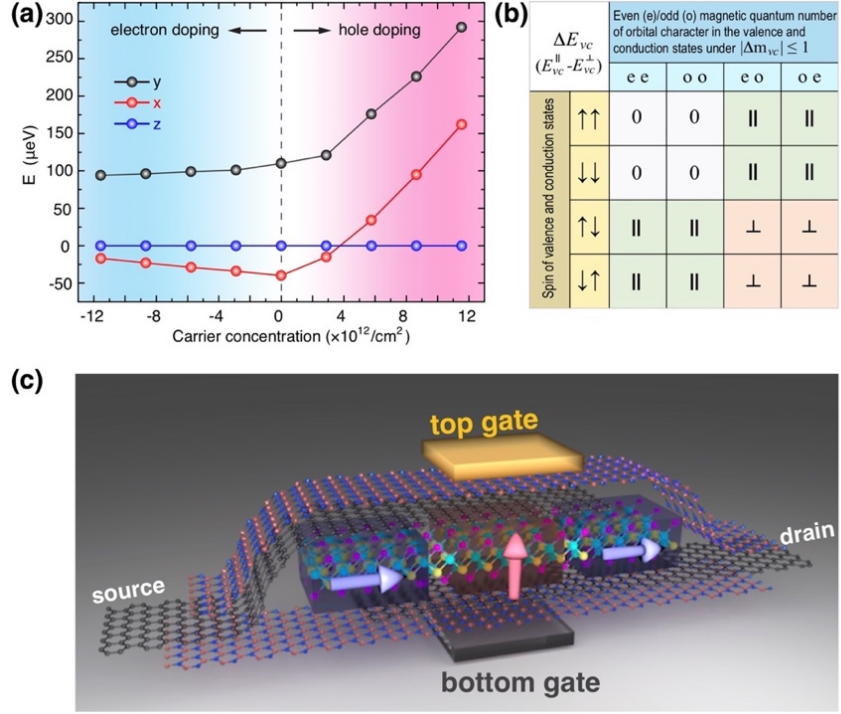

FIG. 3. Electrical control of 2D ferromagnetism in monolayer CrSBr. (a) Energy of FM configurations in different magnetization direction as function of carrier concentration *n*. (b) Contribution of each valence and conduction state pair to the in-plane/out-of-pane magnetization determined by spin and magnetic quantum number of orbital characters in valence and conduction states. (c) A schematic of 2D magnetoelectric device that can realize giant magnetoresistance effect controlled by electrostatic doping.

Doping induced tunability of MAE can be understood from a perturbation theory analysis.[41] Given a pair of valence (*v*) eigenstate $\psi_v$ and conduction (*c*) eigenstate $\psi_c$, their contribution to MAE is given by

$$\Delta E_{vc} = \frac{1}{\Delta_{vc}}(|H_{vc}^{soc}(\vec{x})|^2 - |H_{vc}^{soc}(\vec{z})|^2) \quad (2)$$

where $H_{vc}^{soc}(\vec{n}) = \langle\psi_v|H^{soc}(\vec{n})|\psi_c\rangle$ is the SOC matrix element and $\Delta_{vc} = \varepsilon_v - \varepsilon_c$. $H^{soc}(\vec{n}) = \xi\vec{\sigma}\cdot\vec{L}$, where $\vec{\sigma} = (\sigma_x, \sigma_y, \sigma_z)$ are the 2×2 Pauli matrices, $\vec{L}$ is orbital angular momentum operator, and $\xi$ is the SOC strength. Spin and magnetic quantum number of orbital characters determine the sign of $\Delta E_{vc}$ (Fig. 3(b)), while the sum of $\Delta E_{vc}$ in Eq. 2 over all valence-conduction pairs determines MAE and easy axis. Cr contributes to both in-plane and out-of-plane magnetization, while S and Br favor in-plane easy axis. Due to higher atomic number of Br, its SOC strength is much larger (~3 times of Cr and 30 times of S), hence Br has stronger influence on MAE than Cr and S. Consequently, in-plane easy axis is preferred. Upon electron doping, Cr-$d_{x^2-y^2,\uparrow}$ in the lowest conduction band becomes occupied (Fig. S8), thus $\Delta E_{vc}$ between conduction Cr-$d_{x^2-y^2,\uparrow}$ and valence Cr-$d_{yz,\uparrow}$ favors in-plane magnetization. Increasing occupation in Cr-$d_{x^2-y^2,\uparrow}$ with electron doping reduces the stability of in-plane magnetization. Upon hole doping S-$p_{y,\uparrow}$ and Br-$p_{y,\uparrow}$ in the highest valence band become unoccupied, thereby reducing the stability of in-plane magnetization since $\Delta E_{vc}$ between valence Br-$p_{y,\uparrow}$ and conduction Br-$p_{z,\uparrow}$ favors in-plane magnetization. Due to Br's stronger SOC, hole doping has larger impact on MAE than electron doping, reflected in the stiffer slope upon hole doping (see Fig. 3(a)). Hence, easy axis can be more easily tuned by carrier doping at a critical hole concentration of $4\times10^{12}/cm^2$.

Doping-modulated easy axis allows for realizing 2D spin-FET.[20] A schematic of such magnetoelectric device is proposed in Fig. 3(c), where monolayer CrSBr is double-gated for carrier doping with two dielectric layers (e.g. hexagon BN) to prevent direct tunneling. Gate voltage controls carrier concentration and changes easy axis upon hole doping, while longitudinal in-plane source-drain voltage drives spin-dependent transport. Upon critical hole doping, easy axis switches from in-plane to out-of-plane. Thus, an in-plane FM/out-of-plane FM interface emerges between hole-doped and undoped region. Strong scattering will take place at this hetero-magnetic interface with high resistance, while homo-magnetization below critical doping corresponds to low-resistance state, thereby realizing electrically-controlled giant magnetoresistance effect by dynamically-electrostatic doping.

In summary, we presented a theoretical study of 2D ferromagnetic CrSBr and CrSeBr with high $T_c$ of 168K and 150K and sizable bandgap of 1.66 and 0.78eV, respectively. Remarkably, magnetization easy axis can be tuned from in-plane to out-of-plane by electrostatic doping. Monolayer CrSBr and CrSeBr ferromagnets offer long-desired alternatives to dilute magnetic semiconductors and provide unprecedented opportunities for 2D spintronics such as spin FET.

See supplementary material for details on the calculation methods, crystal and electronic structure, Heisenberg XYZ model of monolayer CrSBr and CrSeBr.

During the submission of this paper, we notice a recent experiment manuscript by Telford et al. which shows AFM-coupled CrSBr layers with individual layer being FM-ordered.[42] Our calculated HSE06 electronic gap of 1.66 eV agrees well with the measured gap of 1.5 eV±0.2 eV. The measured Néel temperature reaches 132K, suggesting $T_c$ of monolayer CrSBr very likely goes beyond 132K.

**ACKNOWLEDGMENT**

H.W. and J.Q. contribute equally to this work. H.W. and X.Q. gratefully acknowledge the support by National Science Foundation (NSF) Grant No. DMR-1753054 and Texas A&M University President's Excellence Fund X-Grants Program and the advanced computing resources provided by Texas A&M High Performance Research Computing. J.Q. acknowledges the financial support from the National Natural Science Foundation of China (Projects No. 11674132 and No. 11974148).

**DATA AVAILABILITY**

The data that support the findings of this study are available from the corresponding author upon reasonable request.